\documentclass[lettersize,journal]{IEEEtran}
\usepackage{enumitem}
\usepackage{booktabs}
\usepackage{multirow}
\usepackage{amsmath,amsfonts}
\usepackage{algorithmic}
\usepackage{algorithm}
\usepackage{array}
\usepackage[caption=false,font=normalsize,labelfont=sf,textfont=sf]{subfig}
\usepackage{textcomp}
\usepackage{stfloats}
\usepackage{url}
\usepackage{verbatim}
\usepackage{graphicx}
\usepackage{cite}
\usepackage{color}

\begin{document}

\title{Physics-informed Neural Networks with Fourier Features for Seismic Wavefield Simulation in Time-Domain Nonsmooth Complex Media}

\author{Yi Ding, Su Chen, Hiroe Miyake, Xiaojun Li
        % <-this % stops a space
\thanks{Manuscript received ****, *** 2024; This work was supported in part by National Natural Science Foundation of China (Grant Nos. 52192675, U1839202), and in part by the China Scholarship Council 202306540044 (Corresponding author: Su Chen.)}
        % <-this % stops a space
\thanks{Yi Ding is with the Key Laboratory of Urban Security and Disaster Engineering of the Ministry of Education, Beijing University of Technology, Beijing 100124, China, and also with the Earthquake Research Institute, The University of Tokyo, Tokyo, 113-0032, Japan (e-mail: dingyi18@mails.ucas.edu.cn).}
        % <-this % stops a space
\thanks{Su Chen and Xiaojun Li are with the Key Laboratory of Urban Security and Disaster Engineering of the Ministry of Education, Beijing University of Technology, Beijing 100124, China, and Xiaojun Li is also with the Institute of Geophysics, China Earthquake Administration, Beijing 100081, China (e-mail: chensuchina@126.com; beerli@vip.sina.com).}
        % <-this % stops a space
\thanks{Hiroe Miyake is with the Earthquake Research Institute, The University of Tokyo, Tokyo, 113-0032, Japan (e-mail: hiroe@eri.u-tokyo.ac.jp).}}

\markboth{Preprint submitted to IEEE}%
{Shell \MakeLowercase{\textit{et al.}}: A Sample Article Using IEEEtran.cls for IEEE Journals}

\IEEEpubid{0000--0000/00\$00.00~\copyright~2024 IEEE}

\maketitle

\begin{abstract}
Physics-informed neural networks (PINNs) have great potential for flexibility and effectiveness in forward modeling and inversion of seismic waves. However, coordinate-based neural networks (NNs) commonly suffer from the "spectral bias" pathology, which greatly limits their ability to model high-frequency wave propagation in sharp and complex media. We propose a unified framework of Fourier feature physics-informed neural networks (FF-PINNs) for solving the time-domain wave equations. The proposed framework combines the stochastic gradient descent (SGD) strategy with an independently pre-trained wave velocity surrogate model to mitigate the singularity at the point source. The performance of the activation functions and gradient descent strategies are discussed through ablation experiments. In addition, we evaluate the accuracy comparison of Fourier feature mappings sampled from different families of distributions (Gaussian, Laplace, and uniform). The second-order paraxial approximation-based boundary conditions are incorporated into the loss function as a soft regularizer to eliminate spurious boundary reflections. Through the non-smooth Marmousi and Overthrust model cases, we emphasized the necessity of the absorbing boundary conditions (ABCs) constraints. The results of a series of numerical experiments demonstrate the accuracy and effectiveness of the proposed method for modeling high-frequency wave propagation in sharp and complex media.
\end{abstract}

\begin{IEEEkeywords}
Physics-informed neural networks (PINNs), Seismic wave propagation simulation, Fourier feature neural networks, Absorbing boundary conditions (ABCs), Spectral bias.
\end{IEEEkeywords}

\section{Introduction}
\IEEEPARstart{I}{mproving} the accuracy and stability of the inner domain and artificial boundaries in the simulation of complex wave propagation problems is a fundamental and common requirement for important inversion techniques \cite{Tarantola_1984_InversionSeismicReflection, Virieux_2009_OverviewFullwaveformInversion, Yang_2023_FWIGANFullWaveformInversion, Song_2023_WeightedEnvelopeCorrelationBased}. For example, full waveform inversion (FWI) is widely used in oil and gas exploration to detect underground structures from artificial earthquake seismograms. By minimizing the differences between observed data and simulated data from the wave equations, FWI can provide high-resolution estimates of subsurface parameters. With the successful application of machine learning in many fields, there is increasing interest in utilizing the approximation capabilities of machine learning techniques for seismic inversion \cite{Yang_2019_DeeplearningInversionNextgeneration, Zhu_2023_FourierDeepONetFourierenhancedDeep}. However, the lack of labeled training datasets is a common challenge in applying deep learning to solve most geoscience problems. Semi-supervised or unsupervised learning stands as a prospective avenue to surmount this obstacle, leveraging prior constraints to formulate unsupervised loss functions for the training of deep neural networks \cite{Wu_2023_SensingPriorConstraints}.

\IEEEpubidadjcol

With the universal approximation theorem of neural networks \cite{TianpingChen_1995_UniversalApproximationNonlinear} and advances in automatic differentiation, deep learning tools are introducing a new trend, offering an alternative approach for solving partial differential equations (PDEs) through the combination of hidden layers to provide nonlinear approximations. The physics-informed neural networks (PINNs) proposed by Raissi et al. \cite{Raissi_2019_PhysicsinformedNeuralNetworks} seamlessly integrate observation and governing physics and train the neural network by minimizing a physics-informed loss function constructed based on PDE residuals and initial/boundary conditions (I/BCs). PINNs provide new perspectives and ideas for addressing the challenges faced by traditional numerical methods in complex domains \cite{Gao_2021_PhyGeoNetPhysicsinformedGeometryadaptive, Diao_2023_SolvingMultimaterialProblems}, and solving inverse problems \cite{Haghighat_2021_PhysicsinformedDeepLearning, Yang_2021_BPINNsBayesianPhysicsinformed}. In the geophysical forward and inverse problems, PINNs have been successfully applied to the Eikonal equation \cite{Smith_2021_EikonetSolvingEikonal, Waheed_2021_PINNeikEikonalSolution, Gou_2023_BayesianPhysicsInformedNeural}, the Maxwell equation \cite{Scheinker_2023_Physicsconstrained3DConvolutional, Piao_2023_DomainadaptivePhysicsinformedNeural}, wave equations for isotropic and anisotropic media in both time \cite{Rasht-Behesht_2022_PhysicsInformedNeuralNetworks, Moseley_2023_FiniteBasisPhysicsinformed, Zhang_2023_SeismicInversionBased, Sethi_2023_HardEnforcementPhysicsinformed}, and frequency domains \cite{Song_2021_SolvingFrequencydomainAcoustic, Song_2021_VersatileFrameworkSolve, Huang_2024_MicroseismicSourceImaging, Alkhalifah_2024_PhysicsinformedNeuralWavefields}.

Current research on PINNs methods for modeling wave propagation behavior has received widespread attention, but there have also been certain challenges in applying vanilla PINNs to seismic wave propagation problems. Firstly, some studies have reported that PINNs suffer from point-source singularity when solving wave equations in the frequency and time domains. Alkhalifah et al. \cite{Alkhalifah_2021_WavefieldSolutionsMachine} addressed the scatter wavefield of the Helmholtz equation based on analytical background wavefields to avoid the singularity arising from high sparsity. To solve the time-domain wave equations, the early initial wavefields obtained by analytical \cite{Zou_2023_NumericalSolverIndependentSeismic} or conventional numerical methods \cite{Moseley_2020_SolvingWaveEquation, Ding_2023_SelfadaptivePhysicsdrivenDeep} can be utilized to provide information about the source. This scheme leads to additional initial conditions loss components, which can be adjusted using a loss weight adaptive algorithm \cite{Wang_2021_UnderstandingMitigatingGradient, Wang_2022_WhenWhyPINNs} to balance the backpropagation gradients between different loss terms.

On the other hand, neural networks tend to learn low-frequency functions and encounter difficulties in learning high-frequency functions, a phenomenon known as "spectral bias" \cite{Rahaman_2019_SpectralBiasNeural}. Tancik et al. \cite{Tancik_2020_FourierFeaturesLet} introduced the Fourier feature network, which employs Fourier feature mapping to transform the input coordinates into a series of high-frequency sinusoidal waves, enhancing the neural network's capability to learn high-frequency functions. Subsequently, the Fourier feature neural networks have been introduced into physics-informed learning \cite{Wang_2021_EigenvectorBiasFourier}, then utilized to solve the multi-frequency Helmholtz equation \cite{Huang_2022_PINNupRobustNeural, Song_2022_SimulatingSeismicMultifrequency} and the acoustic wave equation \cite{Sethi_2023_HardEnforcementPhysicsinformed}.

In conventional discrete numerical algorithms, the discrete equations of motion of the nodes need to be integrated step-by-step using appropriate time integration methods to ensure that the solutions are obtained in temporal order. In contrast, PINNs simultaneously predict the solution over the entire spatio-temporal domain, which may violate temporal causality and tend to converge to erroneous solutions \cite{Krishnapriyan_2021_CharacterizingPossibleFailure, Wang_2024_RespectingCausalityTraining}. The inherent implicit bias in PINNs hinders the learning process associated with long-term complex dynamic processes. A sequential training strategy via time domain decomposition has been introduced to improve the ability of PINNs to model complex wave propagation behavior \cite{Ding_2023_SelfadaptivePhysicsdrivenDeep, Ren_2024_SeismicNetPhysicsinformedNeural}.

Specifying proper initial conditions (ICs) and boundary conditions (BCs) for a PDE is essential to have a well-posed problem. The original formulation of PINNs utilized a “soft constraints” manner to incorporate I/BCs. Emerging studies have rigorously guaranteed the implementation of hard-embedding I/BCs and avoided the issue of imbalanced gradients in multi-component loss functions \cite{Alkhadhr_2023_WaveEquationModeling, Moseley_2023_FiniteBasisPhysicsinformed, Sethi_2023_HardEnforcementPhysicsinformed}. Absorbing boundary conditions (ABCs) are commonly used to model wave propagation in infinite or semi-infinite media under classical numerical methods. At present, a consensus has not been reached regarding the necessity of including ABCs in the solution of the wave equation using PINNs. When addressing the Helmholtz equation in non-smooth media, the incorporation of perfect matching layer conditions into the loss function is proposed to enhance the coupling of the real and imaginary components of the wavefield \cite{Wu_2023_HelmholtzequationSolutionNonsmooth}. Additionally, for modeling seismic wave propagation in 2D elastic media, an absorbing boundary condition is integrated into the network as a soft regularizer to address truncated boundaries \cite{Ren_2024_SeismicNetPhysicsinformedNeural}. However, the potential of PINNs for modeling seismic wavefields in the time domain has not been fully explored, especially concerning complex media and high-frequency wave behavior in the time domain.

Motivated by the foregoing developments, this work proposes a novel unified PINNs architecture for time-domain seismic wavefield simulation in complex media. To alleviate the singularity issue of point sources, we first inject the source using the smoothed Gaussian spatial mapping to approximate the Dirac delta function. Additionally, a fast wave velocity model prediction module is proposed to combine with the stochastic gradient descent (SGD) training strategy, randomly sampling training points at each training step. This allows PINNs to effectively capture information about the seismic source.

To ensure an accurate implementation of the zero-initial condition, the hard embedding scheme proposed by Sethi et al. \cite{Sethi_2023_HardEnforcementPhysicsinformed} is introduced. We carry out a series of ablation experiments to compare the impact of activation functions and training strategies on the proposed framework, as well as the performance of different families of distributions of Fourier feature parameters. Taking the Marmousi and Overthrust models as examples, we show that non-smooth complex medium velocity models lead to inaccurate wavefields when ABCs are not considered. The paraxial approximation technique of the acoustic wave equations is introduced into the loss function as a soft regularization to solve the wave propagation problem in the infinite domain. Combining ABCs and time-domain decomposition training strategies, we demonstrate the potential and versatility of the proposed framework.

\section{Theory}\label{sec:theory}
\subsection{Problem Setting}\label{sec:problem-setting}
In this paper, we aim to investigate the potential of Fourier feature PINNs for modeling wave propagation in complex velocity media. The propagation of seismic waves can be simulated in the time domain by solving the 2D acoustic wave equation as follows:
\begin{equation}\label{eq:1}
\begin{aligned}
\frac{\partial^{2} u(\mathbf{x}, t)}{\partial t^{2}} & = c^{2}(\mathbf{x}) \nabla^{2} u(\mathbf{x}, t)+s(t) G\left(\mathbf{x}, \mathbf{x}_{s}\right) \\
s(t) & = M_{0}\left(1-2\left(\pi f_{0}\left(t-t_{0}\right)\right)^{2}\right) \exp \left(\left(\pi f_{0}\left(t-t_{0}\right)\right)^{2}\right) \\
G\left(\mathbf{x}, \mathbf{x}_{s}\right) & =\exp \left(-\frac{1}{2}\left\|\frac{\mathbf{x}-\mathbf{x}_{s}}{\alpha}\right\|_{2}^{2}\right)
\end{aligned}
\end{equation}
where $\mathbf{x}=\left\{x,z\right\}$ is a vector of spatial coordinates for 2D media, $\mathbf{x}_s=\left\{x_s,z_s\right\}$ denotes the coordinate of the source. $c\left(\mathbf{x}\right)$ is the velocity model and $u\left(\mathbf{x},t\right)$ is the displacement wavefield in the time domain. The source time function $s\left(t\right)$ is injected via a smooth space dependent field $G\left(\mathbf{x},\mathbf{x}_s\right)$ of Gaussian shape. The kernel width $\alpha$ controls the spatial mapping range of the point source, $M_0$ is the amplitude. A Ricker wavelet is used as the source time function, with a dominant frequency of $f_0$ and a delay time of $t_0=\frac{1}{f_0}$.

The initiation of a point-like source at a single grid point can be readily accomplished utilizing the finite differences method. However, this is no longer suitable within the PINNs method. Similar issues are observed with pseudo-spectral methods called the Gibbs phenomenon, as the Fourier transform of a spike-like function creates oscillations that damage the accuracy of the solution \cite{Igel_2017_ComputationalSeismologyPractical}. We use the Gaussian function in (\ref{eq:1}) to define the space-dependent part of the source, injecting it with a smooth spatial distribution of Gaussian shape to alleviate singularity issues associated with point sources.

\subsection{Hard embedding initial conditions in PINNs}\label{sec:hard-embedd-init}
Differing from data-driven neural networks that require a large amount of data to establish the mapping between input and output, the optimization process of PINNs relies only on known physical principles, without the need for any labeled data. Considering the 2D acoustic wave equations described by (1), the residual $\mathcal{R}_r\left(t_r^i,\mathbf{x}_r^i\right)$ and loss function $\mathcal{L}_r\left(\boldsymbol{\theta}\right)$ corresponding to the PDEs are

\begin{align}\label{eq:2}
\mathcal{R}_{r}\left(t_{r}, \mathbf{x}_{r}\right) &=\frac{\partial^{2} \hat{u}_{\boldsymbol{\theta}}}{\partial t_{r}^{2}}-c^{2} \nabla^{2} \hat{u}_{\boldsymbol{\theta}}+s\left(t_{r}\right) G\left(\mathbf{x}_{r}, \mathbf{x}_{s}\right), \\\label{eq:3}
\mathcal{L}_{r}(\boldsymbol{\theta}) &=\frac{1}{N_{r}} \sum_{i=1}^{N_{r}}\left|\mathcal{R}_{r}\left(t_{r}^{i}, \mathbf{x}_{r}^{i}\right)\right|^{2},
\end{align}
where ${\hat{u}}_{\boldsymbol{\theta}}$ is the unknown potential solution approximated by the deep neural networks. $\left\{t_r^i,\ \mathbf{x}_r^i\right\}_{i=1}^{N_r}$ denotes the collocation points of the PDEs residual, with the number being $N_r$.

\begin{figure*}[!t]
\centering
\includegraphics[width=5.5in]{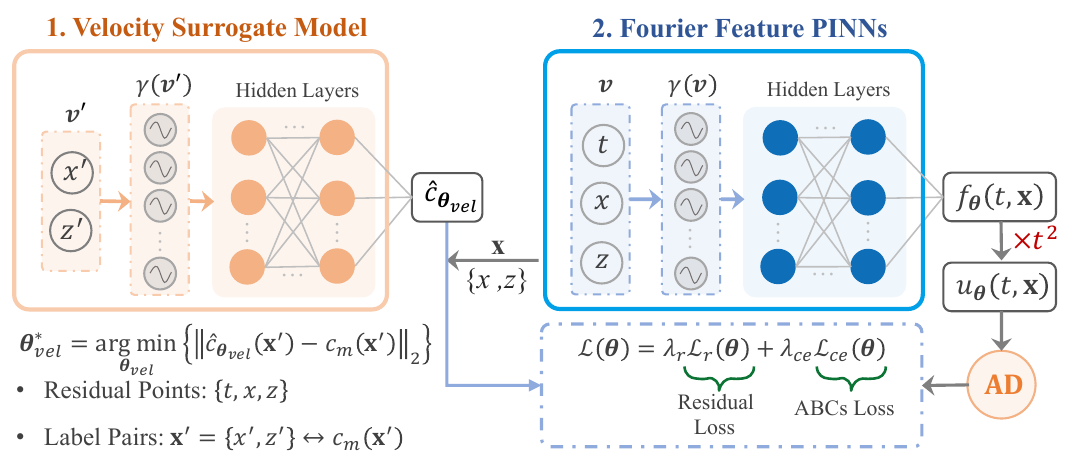}
\caption{Schematic diagram of the proposed framework. 1. Train a wave velocity surrogate model using the coordinate-velocity label pairs ($\mathbf{x}^{\prime}=\left\{x^{\prime},z^{\prime}\right\} \leftrightarrow c_m(\mathbf{x}^{\prime})$) of the computational model. 2. FF-PINNs is used to model wave propagation. Using the SGD training strategy means randomly sampling the collocation points ($\{t, \mathbf{x}\}$) at each epoch, predicting the wave velocity $\hat{c}_{\boldsymbol{\theta}_{vel}}(\mathbf{x})$ at each sampling point using the trained wave velocity surrogate model, and constructing loss functions ($\mathcal{L}_r(\boldsymbol{\theta})$ and $\mathcal{L}_{ce}(\boldsymbol{\theta})$) for optimization. AD in the figure denotes automatic differentiation.}
\label{fig_framework}
\end{figure*}

In the vanilla PINNs framework, I/BCs are imposed as loss terms in the target function in a "soft constraint" manner. The gradient imbalance problem of different loss terms may lead to the failure of neural network training, and the weight parameters of different loss terms must be carefully determined. Meanwhile, the optimization process cannot guarantee that the soft constraint loss term is completely zero. The output of NNs is directly the displacement solution of the second order wave equations, which can directly satisfy the initial and boundary conditions related to $u$. To ensure a unique solution to the wave equations, it is imperative to rigorously define the initial conditions. Before the seismic source excitation, the system is in a quiescent state with zero initial conditions. The neural network output $f_{\boldsymbol{\theta}}\left(t,\mathbf{x}\right)$ can be transformed as
\begin{equation}\label{eq:4}
\hat{u}_{\boldsymbol{\theta}}(t, \mathbf{x})=f_{\boldsymbol{\theta}}(t, \mathbf{x}) \cdot t^{2},
\end{equation}
which guarantees that ${\hat{u}}_{\boldsymbol{\theta}}=\partial{\hat{u}}_{\boldsymbol{\theta}} / \partial t=0$ at $t=0$, thus realizing hard embedding with zero initial conditions. Here $t^2$ is used as the approximate distance function (ADF) and is not unique. The employed ADF must satisfy two required criteria: 1) The ADF must be equal to zero at the initial moment $t=0$, and 2) for any $t>0$, the ADF and its gradient must be smooth and non-zero. Fig. \ref{fig_framework} illustrates the structure of the proposed framework.

\subsection{Fourier Feature Mapping}\label{sec:four-feat-mapp}
With the Neural Tangent Kernel (NTK) theory \cite{Jacot_2018_NeuralTangentKernel}, the "spectral bias" pathology of neural networks can be rigorously analyzed and elucidated. The neural network is biased to first learn the target function along the eigendirections of NTK with larger eigenvalues, and then learn the remaining components corresponding to the smaller eigenvalues \cite{Wang_2021_EigenvectorBiasFourier}. However, NTK theory suggests that this is because standard coordinate-based multilayer perceptrons (MLPs) correspond to kernels with a rapid frequency falloff, which effectively prevents them from learning the high-frequency functions \cite{Rahaman_2019_SpectralBiasNeural, Ronen_2019_ConvergenceRateNeural, Tancik_2020_FourierFeaturesLet}.

In PINNs methods, MLPs are generally used as universal approximators to learn the solutions of PDEs. Let $\boldsymbol{v}=\left\{t,\mathbf{x}\right\}\in\mathbb{R}^d$ denote the input coordinates. The Fourier feature NNs accept the coordinates as input and map them to a feature space with a Fourier feature mapping before passing them to the hidden layer. Following the original formulation of Tancik et al. \cite{Tancik_2020_FourierFeaturesLet}, the Fourier feature neural network is
\begin{equation}\label{eq:5}
\begin{aligned}
\gamma(\boldsymbol{v}) & =\left[\begin{array}{l}
\cos (2 \pi \mathbf{B} \boldsymbol{v}) \\
\sin (2 \pi \mathbf{B} \boldsymbol{v})
\end{array}\right], \\
\boldsymbol{H}^{(1)} & =\phi\left(\boldsymbol{W}^{(1)} \cdot \gamma(\boldsymbol{v})+\boldsymbol{b}^{(1)}\right), \\
\boldsymbol{H}^{(l)} & =\phi\left(\boldsymbol{W}^{(l)} \cdot \boldsymbol{H}^{(l-1)}+\boldsymbol{b}^{(l)}\right), \quad l=2, \cdots, L, \\
\boldsymbol{f}_{\boldsymbol{\theta}}(\boldsymbol{v}) & =\boldsymbol{W}^{(L+1)} \cdot \boldsymbol{H}^{(L)}+\boldsymbol{b}^{(l)},
\end{aligned}
\end{equation}
where $\boldsymbol{W}^{\left(l\right)}\in\mathbb{R}^{d_l\times d_{l-1}}$, $\mathbf{b}^{\left(l\right)}\in\mathbb{R}^{d_l}$ are the trainable weight and bias of the lth layer, respectively. $\phi$ denotes the nonlinear activation function. Fourier basis frequency $\mathbf{B}\in\mathbb{R}^{m\times d}$ can be sampled from different distribution families (Gaussian, Laplacian, uniform), leading to different ways of embedding Fourier features. For example, in a Gaussian Fourier feature NNs, each entry of $\mathbf{B}$ is i.i.d. sampled from a normal distribution $\mathcal{N}\left(0,\sigma^2\right)$ with the standard deviation $\sigma>0$. $m$ is the number of basis functions in the Fourier basis set. During the network training process, $\mathbf{B}$ can be kept constant or optimized alongside the weights and biases as the trainable parameter $\boldsymbol{\theta}$. In all the cases considered in this work, we set $\mathbf{B}$ as a fixed parameter and let $m=256$.

The standard deviation $\sigma$ is a hyperparameter that directly governs the scale of the distribution of the Fourier base frequencies and the resulting eigenspace of the NTK. It is usually specified empirically or requires an expensive hyperparameter search to determine. The performance of FF-PINNs is strongly influenced by the hyperparameter $\sigma$, and optimal performance is only achieved when $\sigma$ is tuned to a certain range. For specific problems, prior frequency information can be obtained from available I/BCs, source time functions, or observations to avoid extensive and expensive hyperparameter searches.

\subsection{Absorbing boundary conditions constraints}\label{sec:absorb-bound-cond}
ABCs play a crucial role in the numerical simulation of wave propagation problems. To achieve the goal that all outgoing waves are not reflected by boundaries, the computed boundary motions should take into account all the complexities of the outgoing waves. The paraxial boundary conditions are derived by rational approximation of the 2D acoustic-wave dispersion relation \cite{Clayton_1977_AbsorbingBoundaryConditions, Engquist_1977_AbsorbingBoundaryConditionsa, Xing_2021_TheoryNewUnified}. For example, $ck_z / \omega=-\sqrt{1-s^2}$ with respect to the variable $s=ck_x=\omega$ in the negative $z$ direction. Thus, the paraxial approximation technique can be strictly used for constructing ABCs for 2D acoustic wave equations.

In PINNs, it is straightforward to impose paraxial-approximation-based ABCs at the boundary in a soft constraint manner. ABCs in continuous form are treated directly in PINNs, the relevant derivatives are computed by automatic differentiation without the need for numerical discretization. In this study, a second-order expression of the Clayton-Engquist acoustic paraxial approximation boundary \cite{Clayton_1977_AbsorbingBoundaryConditions} is considered. The computational model is a rectangular region, the coordinate origin is located in the upper left corner of the model, and the boundary directions are parallel to \(x\) and $z$ coordinate axes, respectively. Then residuals of the second-order Clayton-Engquist boundary conditions can be expressed as

\noindent $x$-axis positive:
\begin{equation}\label{eq:6}
\mathcal{R}_{x l}=\frac{\partial^{2} \hat{u}_{\boldsymbol{\theta}}}{\partial x_{l} \partial t_{c}}+\frac{1}{c} \frac{\partial^{2} \hat{u}_{\boldsymbol{\theta}}}{\partial t_{c}^{2}}-\frac{c}{2} \frac{\partial^{2} \hat{u}_{\boldsymbol{\theta}}}{\partial z_{c}^{2}},
\end{equation}
$x$-axis negative:
\begin{equation}\label{eq:7}
\mathcal{R}_{x b}=\frac{\partial^{2} \hat{u}_{\boldsymbol{\theta}}}{\partial x_{b} \partial t_{c}}-\frac{1}{c} \frac{\partial^{2} \hat{u}_{\boldsymbol{\theta}}}{\partial t_{c}^{2}}+\frac{c}{2} \frac{\partial^{2} \hat{u}_{\boldsymbol{\theta}}}{\partial z_{c}^{2}},
\end{equation}
$z$-axis positive:
\begin{equation}\label{eq:8}
\mathcal{R}_{z l}=\frac{\partial^{2} \hat{u}_{\boldsymbol{\theta}}}{\partial z_{l} \partial t_{c}}+\frac{1}{c} \frac{\partial^{2} \hat{u}_{\boldsymbol{\theta}}}{\partial t_{c}^{2}}-\frac{c}{2} \frac{\partial^{2} \hat{u}_{\boldsymbol{\theta}}}{\partial x_{c}^{2}},
\end{equation}
$z$-axis negative:
\begin{equation}\label{eq:9}
\mathcal{R}_{z b}=\frac{\partial^{2} \hat{u}_{\boldsymbol{\theta}}}{\partial z_{b} \partial t_{c}}-\frac{1}{c} \frac{\partial^{2} \hat{u}_{\boldsymbol{\theta}}}{\partial t_{c}^{2}}+\frac{c}{2} \frac{\partial^{2} \hat{u}_{\boldsymbol{\theta}}}{\partial x_{c}^{2}} .
\end{equation}

\begin{figure}[!t]
\centering
\includegraphics[width=1.8in]{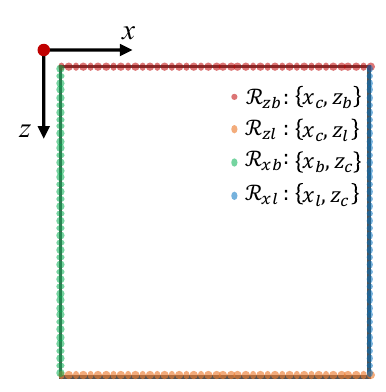}
\caption{The distribution of the applied ABCs collocation points along the four boundaries of the rectangular domain.}
\label{fig_ABCs}
\end{figure}

The loss function now consists of the PDEs residual loss defined in (3) and the ABCs residual loss $\mathcal{L}_{ce}$,
\begin{equation}\label{eq:10}
\begin{aligned}
\mathcal{L}(\boldsymbol{\theta}) & =\lambda_{r} \mathcal{L}_{r}(\boldsymbol{\theta})+\lambda_{c e} \mathcal{L}_{c e}(\boldsymbol{\theta})=\frac{\lambda_{r}}{N_{r}} \sum_{i=1}^{N_{r}}\left|\mathcal{R}_{r}\right|^{2}  \\
& +\frac{\lambda_{c e}}{N_{c e}} \sum_{i=1}^{N_{c e}}\left (\left|\mathcal{R}_{x l}\right|^{2}+\left|\mathcal{R}_{x b}\right|^{2}+\left|\mathcal{R}_{z l}\right|^{2}+\left|\mathcal{R}_{z b}\right|^{2}\right)
\end{aligned}
\end{equation}
where $\left\{x_b,z_b,x_l,z_l\right\}$ denotes the collocation points from the spatial boundaries with the number being $N_{ce}$, the subscript $b$ is the lower bound and $l$ is the upper bound. $x_c$, $z_c$ and $t_c$ are randomly sampled collocation points across the $x$, $z$ and $t$ axes, respectively. Fig. \ref{fig_ABCs} illustrates the distribution of boundary collocation points for a rectangular domain with ABCs imposed. $\lambda_r$ and $\lambda_{ce}$ are the weight hyperparameters of the corresponding loss function terms. Algorithm \ref{alg1} demonstrates the training pipeline of the proposed framework.

\begin{algorithm}[!t]
\caption{Training pipeline of the proposed framework}\label{alg1}
\begin{algorithmic}
\STATE\begin{enumerate}[nolistsep]
\item Initializing MLPs ${\hat{c}}_{\boldsymbol{\theta}_{vel}}$ and ${\hat{u}}_{\boldsymbol{\theta}}$ with Fourier feature mapping using the Glorot scheme.
\item Optimize the velocity NN ${\hat{c}}_{\boldsymbol{\theta}_{vel}}$ using coordinate-velocity label pairs with the Adam algorithm.
\item Hard embedding zero initial conditions with (\ref{eq:4}).
\item Optimize the parameters $\boldsymbol{\theta}$ with an SGD algorithm within $N_{iter}$ epochs: \label{item:1}
\end{enumerate}
\FOR{$n=1,\cdots, N_{iter}$}
\IF{w/ ABCs}
\STATE Randomly sample $\left\{t_r, \mathbf{x}_r\right\}$.
\STATE Predict ${\hat{c}}_{\boldsymbol{\theta}_{vel}}$ at $\mathbf{x}_r$.
\STATE Calculate $\mathcal{L}_r\left(\boldsymbol{\theta}\right)$ based on (\ref{eq:3}).
\ELSIF{w/o ABCs}
\STATE Randomly sample $\left\{t_r,\mathbf{x}_r\right\}$ and $\left\{t_{ce},\mathbf{x}_{ce}\right\}$.
\STATE Predict ${\hat{c}}_{\boldsymbol{\theta}_{vel}}$ at $\mathbf{x}_r$ and $\mathbf{x}_{ce}$.
\STATE Calculate $\mathcal{L}_r\left(\boldsymbol{\theta}\right),\mathcal{L}_{ce}\left(\boldsymbol{\theta}\right)$ based on (\ref{eq:10}).
\ENDIF
\STATE Update the parameters $\boldsymbol{\theta}$ via gradient descent:
\STATE $\boldsymbol{\theta}_{n+1}=\boldsymbol{\theta}_{n}-\eta \nabla_{\boldsymbol{\theta}} \mathcal{L}\left(\boldsymbol{\theta}_{n}\right) $
\ENDFOR
\STATE If use time domain decomposition algorithm, change the intervals $\left[0,\ t_i\right]\left(i\ =1,\ 2,\ \ldots\ \text{and}\ 0<t_1<t_2<\cdots T\right)$ and repeat step \ref{item:1}).
\end{algorithmic}
\end{algorithm}

\subsection{Experimental Setup}\label{sec:experimental-setup}
\textbf{Activation functions}. The choice of activation functions has a significant effect on the training dynamics and task performance. Popular activation Rectified Linear Unit (ReLU) is deficient for high-order PDEs since its second-order derivative is zero. The utilization of the hyperbolic tangent (Tanh) activation function has been prevalent in previous research pertaining to PINNs. However, it has been observed that Tanh encounters challenges when tasked with learning high-frequency and periodic functions \cite{Sitzmann_2020_ImplicitNeuralRepresentations, Wang_2024_MultistageNeuralNetworks}. In the field of Implicit Neural Representation (INR), various activation functions have been suggested to help coordinate-MLPs encode high-frequency signals, including periodic (sinusoidal \cite{Sitzmann_2020_ImplicitNeuralRepresentations}) and nonperiodic (e.g., Gaussian \cite{Ramasinghe_2022_PeriodicityUnifyingFramework}) functions. In addition, Swish \cite{Ramachandran_2017_SwishSelfGatedActivation}, an activation function discovered by combining exhaustive and reinforcement learning-based search, achieves impressive performance in many challenging tasks. We carry out a detailed experimental comparison in Section \ref{sec:ablation-study} to provide guidance on the choice of activation functions.

\textbf{Random sampling}. In comparison to batch gradient descent (BGD), stochastic gradient descent (SGD) \cite{Bottou_2018_OptimizationMethodsLargeScale} significantly reduces the memory requirements and the computational cost of each iteration. Moreover, according to our observation, the use of random sampling plays an important role in the training efficiency and model performance, whereas the use of full-batch gradient descent to train PINNs may lead to overfitting of PDE residuals. The continuously changing random sampling across the entire domain ensures that the neural network gains sufficient understanding of the point source region, even if it is sparse relative to the entire spatial domain, enabling PINNs to fully capture information about the seismic source. Section \ref{sec:four-feat-veloc} introduces the velocity surrogate model combined with SGD algorithm. A detailed comparison of the performance between SGD and BGD algorithms will be conducted in the ablation study in Section \ref{sec:ablation-study}.

\textbf{Optimizer}. In this study, only Adam \cite{Kingma_2015_AdamMethodStochastic} was used as the default optimizer, random sampling from the spatio-temporal domain at each epoch. We set the initial learning rate to $5 \times 10^{-3}$ and an exponential decay with a decay rate of 0.9 for every 1000 decay epochs. Since the L-BFGS \cite{Liu_1989_LimitedMemoryBFGS} strongly relies on the historical gradient to approximate the inverse Hessian matrix, when these gradients are computed using collocation points the process can be unstable. The stable implementation of quasi-Newton updating in a multi-batch setting continues to garner significant attention in the academic literature \cite{Berahas_2016_MultiBatchLBFGSMethod, Berahas_2020_RobustMultibatchLBFGS}. Finally, dense layers will be typically initialized using the Glorot scheme \cite{Glorot_2010_UnderstandingDifficultyTraining}.

\textbf{Time domain decomposition strategy}. In this study, we have employed a time-domain decomposition strategy to address the wave propagation problem in complex media. The time-domain decomposition decomposes the difficult optimization problem over the whole time period into several relatively simple problems, which significantly reduces the optimization difficulty of learning the complete evolution of the dynamical system. Specifically, if the entire simulation time domain is $\left[0,\ T\right]$, the solution of PDEs is sequentially learned in the time period $\left[0,\ t_i\right]\left(i\ =1,\ 2,\ \ldots\ \text{and} \ 0<t_1<t_2<\cdots T\right)$. The training parameters of the previous time period are passed to the next time period to provide good parameter initialization.

\textbf{Evaluation}. Predictions ${\hat{u}}_{\boldsymbol{\theta}}$ for PINNs/FF-PINNs and ${\hat{c}}_{\boldsymbol{\theta}_{vel}}$ for the wave velocity surrogate model are evaluated using the relative $\ell_2$ error
\begin{equation}\label{eq:11}
\epsilon_{\ell_{2}}=\sqrt{\frac{\sum_{i=1}^{n}\left(\hat{y}_{pred}-y_{ref}\right)}{\sum_{i=1}^{n} y_{ref}}}
\end{equation}
where ${\hat{y}}_{pred}$ is the prediction result of the neural networks, and $y_{ref}$ represents the ground truth. $n$ is the number of test sampling points.

In order to assess the accuracy of the proposed method for solving the wave equation, the results of the finite difference method (FDM) are used as reference data. We use unsplit convolutional perfectly matched layer (PML) boundary condition \cite{Komatitsch_2007_UnsplitConvolutionalPerfectly} for staggered-grid second-order finite difference modeling of the time-dependent 2D acoustic wave equation. For the infinite medium wave propagation problem, we add 10-grid PML boundaries to each of the four boundaries of the rectangular domain to prevent spurious boundary-reflected waves.

\subsection{The Fourier feature velocity surrogate model}\label{sec:four-feat-veloc}
Previous studies have avoided the tendency of neural networks to converge to a zero solution by ensuring that sufficient collocation points are added near the source location. However, this approach results in a significant amount of computation and may lead to erroneous attention mechanisms during the training process, as the NNs excessively focus on optimizing around the source while neglecting other areas. According to our experiments, random sampling in the spatio-temporal domain in each iteration and applying SGD can sufficiently learn the information of the source, thus avoiding additional sampling in the region around the source.

Due to the heterogeneity of the model, the wave velocity corresponding to the randomly sampled spatial coordinates must be obtained quickly in each epoch to participate in the calculation of the PDE residual loss. To solve this problem, we use a supervised neural network for fast prediction of wave velocity models. The velocity model prediction network is trained using coordinate-velocity label pairs ($\mathbf{x}^{\prime}=\left\{x^{\prime},z^{\prime}\right\} \leftrightarrow c_m(\mathbf{x}^{\prime})$) before the simulation starts. The input to the network is the spatial coordinates $\{\mathbf{x}^{\prime}\}$ and the output is the corresponding wave velocity $\hat{c}_{\boldsymbol{\theta}_{vel}}$. The training of the supervised neural network can be viewed as solving an optimization problem described as

\begin{equation}
\boldsymbol{\theta}^*_{vel}=\arg\min_{\boldsymbol{\theta}_{vel}} \{\left\|\hat{c}_{\boldsymbol{\theta}_{vel}}(\mathbf{x}^{\prime}) - c_m(\mathbf{x}^{\prime})\right\|_2\}
\end{equation}

Upon completion of the velocity network training, the prediction of velocity can be executed in each subsequent PINNs/FF-PINNs training epoch, rendering the impact on training time negligible. The predictive efficiency and accuracy of the velocity model neural network are crucial to the overall workflow. Since neural networks tend to learn low-frequency functions, a reasonable conjecture is that they may struggle to learn the detailed structure of complex velocity models. We use a comparative experiment with the Overthrust velocity model to demonstrate the importance of Fourier feature mapping. The Fourier feature NNs and fully connected NNs followed the same hyperparameter settings. We trained 5-layer neural networks with 20 neurons per hidden layer for 100,000 epochs using the Adam optimizer, the activation function used is Swish. Scaling of the wave velocity and spatial coordinates by the same proportion was implemented according to the principle of wave velocity normalization.

\begin{table*}[!t]
  \caption{The Relative $\ell_2$ Errors of Predicted Wave Velocity for Fully Connected NNs and Fourier Feature NNs with Different $\sigma$}
  \label{tab:1}
  \centering
\begin{tabular}{ccccccc}
  \toprule
  \multirow{2}{*}{ Fully connected NNs } & \multicolumn{6}{c}{ Fourier feature NNs } \\
\cmidrule { 2 - 7 } & \( \sigma=1 \) & \( \sigma=5 \) & \( \sigma=10 \) & \( \sigma=15 \) & \( \sigma=20 \) & \( \sigma=25 \) \\
  \midrule \( 2.166 \times 10^{-2} \) & \( 4.307 \times 10^{-3} \) & \( 1.994 \times 10^{-3} \) & \( 1.520 \times 10^{-3} \) & \( 1.483 \times 10^{-3} \) & \( 1.746 \times 10^{-3} \) & \( 1.917 \times 10^{-3} \) \\
    \bottomrule \\
\end{tabular}
\end{table*}

\begin{figure}[!t]
\centering
\includegraphics[width=3.2in]{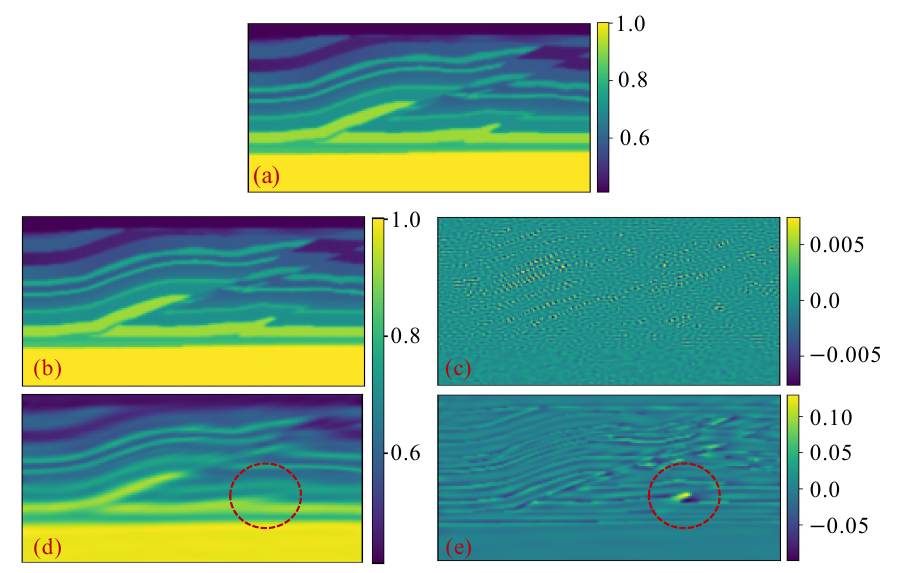}
\caption{Comparison of normalized wave velocity predictions for Fourier feature NNs and fully connected NNs. (a) Ground Truth. (b) Prediction of Fourier feature NNs with $\sigma=15$. (c) Absolute error in prediction of Fourier feature NNs with $\sigma=15$. (d) Prediction of the fully connected NNs. (e) Absolute error in prediction of fully connected NNs.}
\label{fig_velocity_surrogate}
\end{figure}

Table \ref{tab:1} shows the comparison of the relative $\ell_2$ errors in the predictions of fully connected NNs and Fourier feature NNs with different $\sigma$. As can be seen from Table \ref{tab:1}, the using of Fourier feature mapping improves the prediction accuracy by about an order of magnitude. The relative $\ell_2$ error of the prediction of the fully connected NNs is $2.166 \times 10^{-2}$, while that of the Fourier feature NNs with $\sigma=15$ is $1.483 \times 10^{-3}$. A comparison of the absolute errors of the predictions of fully connected NNs and $\sigma=15$ Fourier feature NNs with the true velocity model is shown in Fig. \ref{fig_velocity_surrogate}. For complex velocity models with sharp interfaces, fully connected NNs are unable to learn high-frequency information and thus give blurred predictions because of spectral bias pathology. On the other hand, the Fourier feature NNs showed low sensitivity to $\sigma$ in the tests, achieving low errors (${\epsilon_\ell}_2<2 \times 10^{-3}$) over a wide range ($\sigma\in\left\{5,\ 25\right\}$). The Fourier feature velocity surrogate model proposed in this section provides a robust and accurate prediction of the medium model for the next stage of the solution process.

\section{Ablation Study}\label{sec:ablation-study}
We conduct a detailed ablation study through a homogeneous medium case, which is used to evaluate several aspects that researchers and practitioners should consider when solving the wave equation using FF-PINNs. The discussion highlights the importance of choosing appropriate activation functions, random sampling strategies, and scaling of Fourier feature mappings.

The region of the computational model is $x\in\left[0,\ 600\right]$ m, $z\in\left[0,\ 600\right]$ m, with $c=500$ m/s. The total computation time is 0.9 s. Since the velocity spatial independence in the case of homogeneous media, it can be explicitly defined during the iterative process, obviating the need for the Fourier feature velocity surrogate model mentioned in Section \ref{sec:four-feat-veloc}. We do not consider ABCs in this case, thus the loss function contains only the PDE loss term. In the vanilla PINNs method and FF-PINNs, a fully connected neural network with 5 hidden layers and 50 neurons per layer was used. The number of collocation points for the PDEs residual is $N_r=3,0000$. The results are averaged from 5 independent trials optimized by the Adam optimizer for 10, 000 epochs.

\subsection{Activation functions and random sampling strategies}\label{sec:activ-funct-rand}
\begin{figure}[!t]
\centering
\includegraphics[width=3.2in]{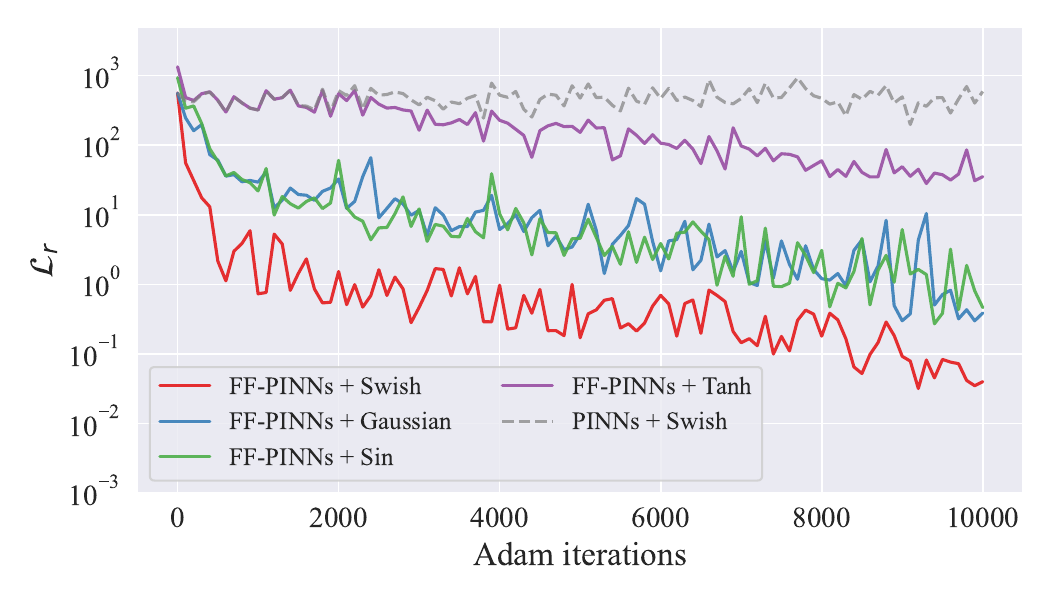}
\caption{The PDE residual loss $\mathcal{L}_r$ evolution using FF-PINNs with different activation functions and PINNs with Swish activation function. Recorded every 100 Adam epochs.}
\label{fig_loss_plot}
\end{figure}

\begin{figure}[!t]
\centering
\includegraphics[width=3.2in]{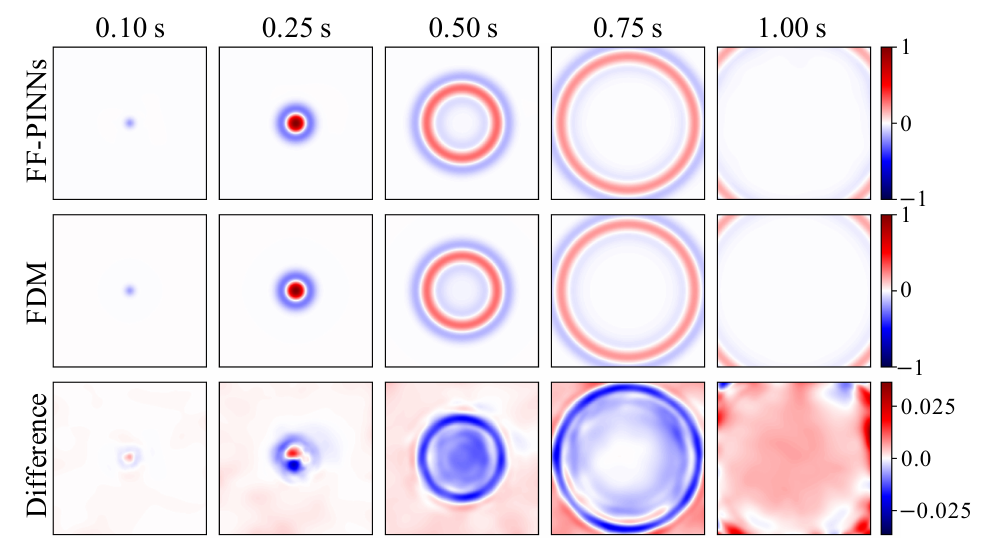}
\caption{Comparison of wavefields of FF-PINNs ($\sigma=1$) and FDM in the homogeneous medium case. SGD training strategy as well as Swish activation function are used.}
\label{fig_homo_wavefield}
\end{figure}

The central frequency of the Ricker source time function considered is $f_0=10$ Hz, and $\alpha=0.02$. We used the SGD algorithm to evaluate the performance of some of the activation functions discussed in Section \ref{sec:experimental-setup}, and show all the results in Table \ref{tab:2}. The commonly used Tanh activation function performs poorly (${\epsilon_\ell}_2=0.6212$) even when $\sigma$ is set to the proper value. In addition, Swish outperforms Gaussian and Sin in all cases, with Gaussian slightly outperforming Sin. The best performance (${\epsilon_\ell}_2=0.0398$) was achieved with Swish activation when $\sigma=1$ for the Gaussian Fourier feature network. As discussed in Section \ref{sec:experimental-setup}, the BGD strategy cannot be used in this problem, probably due to insufficient learning of information about the point sources. The results in Table \ref{tab:2} and Fig. \ref{fig_loss_plot} validate this finding. It is worth noting that embedding Fourier features leads to a slight increase in wall-clock time. In this case, PINNs took 3 minutes to train 10,000 epochs with the Adam optimizer, while FF-PINNs took 4.3 minutes. However, the FF-PINNs obtain a huge performance improvement compared to the PINNs, as can be seen from the evolution of the loss curves in Fig. \ref{fig_loss_plot}. A comparison of the wavefield snapshots of the FF-PINNs prediction with the FDM results for five moments is shown in Fig. \ref{fig_homo_wavefield}. All subsequent experiments herein employ Swish activation and SGD training strategy.

\subsection{Fourier mapping}\label{sec:fourier-mapping}
The parameter $\mathbf{B}$ determines the preferred frequency range learned by the Fourier feature network described by (5), and thus the reasonable selection of the distribution of $\mathbf{B}$ is a prerequisite for accurate prediction by FF-PINNs. In Gaussian Fourier feature neural networks, the scale factor $\sigma$ controls the range of the $\mathbf{B}$ distribution. Previous works that utilized FF-PINNs to simulate the wave equation in the frequency \cite{Song_2022_SimulatingSeismicMultifrequency} and time domains \cite{Sethi_2023_HardEnforcementPhysicsinformed} sampled $\mathbf{B}$ from a uniform distribution with zero as the axis of symmetry. We use Fourier feature mappings sampled from different distribution families (Gaussian, Laplacian, and uniform) and sweep over the standard deviation of each distribution. The Gaussian distribution is $\frac{1}{\sqrt{2\pi}\sigma}e^{-\frac{\left(x-\mu\right)^2}{2\sigma^2}}$, where $\sigma$ is the standard deviation. The Laplacian distribution is $\frac{1}{2b}e^{-\frac{\left|x-\mu\right|}{b}}$, where $b>0$ is a scale parameter, which is sometimes referred to as the ``diversity''. In the Laplacian distribution, the standard deviation $\sigma=\sqrt2b$. The standard deviation of a uniform distribution with distribution range $\left[-b_{max},b_{max}\right]$ is $b_{max} / \sqrt3$. Building on the previous section, a source with $f_0=10$ Hz, $\alpha=0.02$ was considered in this section.

\begin{figure}[!t]
\centering
\includegraphics[width=3.0in]{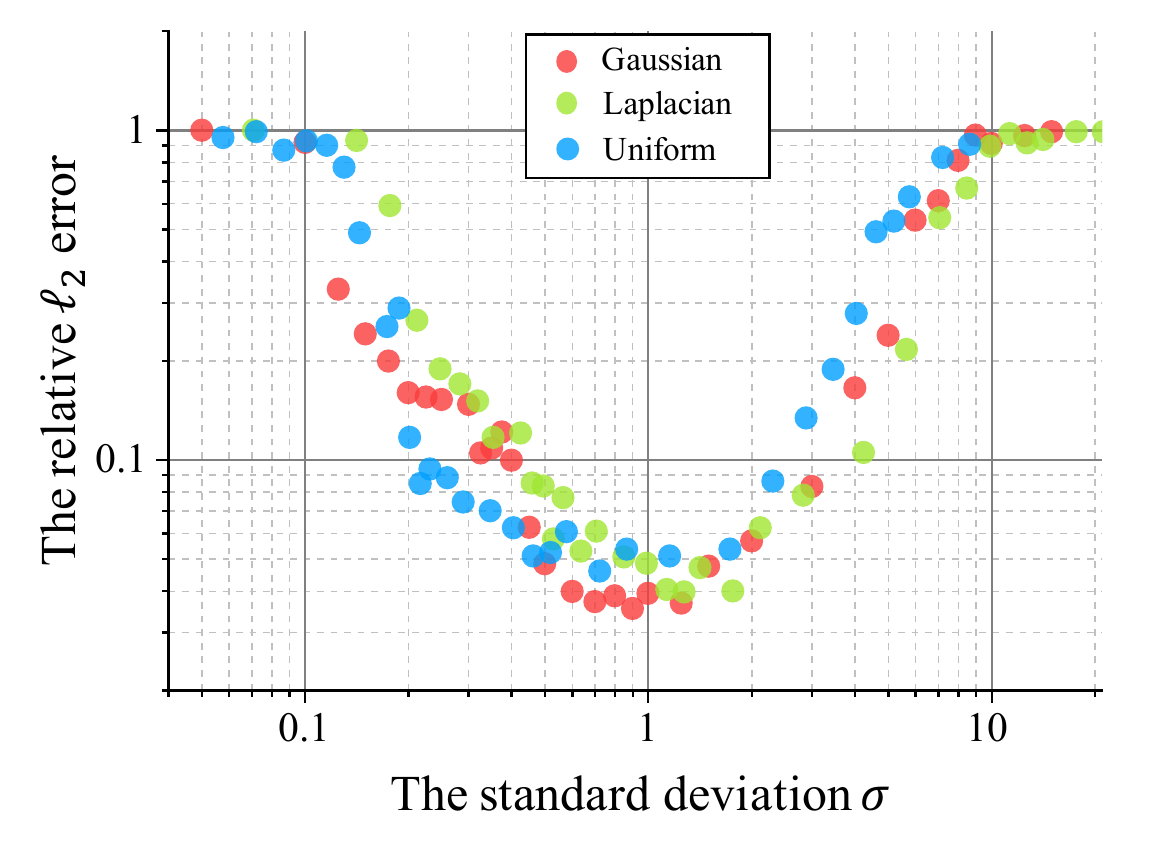}
\caption{Relative $\ell_2$ error results for $\sigma$ hyperparameter sweeps with Fourier features sampled from different distribution families.}
\label{fig_FF_distribution_families}
\end{figure}

Fig. \ref{fig_FF_distribution_families} shows the results of hyperparameter sweeps. The scatter plots of the results induced by three distribution families exhibit a characteristic "U-shaped" pattern. Lower $\sigma$ may make it difficult to learn information about point sources, while higher coding frequencies can introduce salt-and-pepper artifacts. Consequently, both excessively small and large values result in comparatively large prediction errors. Ideally, an appropriate $\sigma$ should be selected such that the bandwidth of NTK matches that of the target signals. This not only accelerates the training convergence but also improves the prediction accuracy. However, the spectral information of the solution may not be accessible when solving forward PDEs. In this work, we perform a proper hyperparameter search starting from $\sigma=1$. From the ablation study in this section, it can be observed that the exact sampling distribution family is much less important than the distribution’s scale (standard deviation), consistent with the findings of Tancik et al. \cite{Tancik_2020_FourierFeaturesLet} in their experiments on 1D target signals. In the subsequent experiments, we only use Gaussian Fourier features PINNs to model the wave forward propagation.

\begin{table*}[!t]
  \caption{The Relative $\ell_2$ Errors of Different Activation Functions and Gradient Descent Strategies}
  \label{tab:2}
\centering
\begin{tabular}{cccccccc}
  \toprule[1pt]
  \multirow{2}{*}{ Algorithm } & \multirow{2}{*}{ Activation } & \multirow{2}{*}{ Equation* } & \multirow{2}{*}{ PINNs } & \multicolumn{4}{c}{ Gaussian Fourier feature PINNs } \\
\cmidrule{5-8} & & & & \( \sigma=0.5 \) & \( \sigma=1 \) & \( \sigma=2.5 \) & \( \sigma=5 \) \\
  \midrule
  \multirow{4}{*}{ SGD } & Tanh & \( \frac{e^{x}-e^{-x}}{e^{x}+e^{-x}} \) & 0.9991 & 0.9996 & 0.6212 & 0.8567 & 0.9993 \\
& Sin & \( \sin (a x) \) & 0.9990 & 0.1970 & 0.1074 & 0.2946 & 0.8208 \\
& Gaussian & \( e^{-\frac{x^{2}}{2 a^{2}}} \) & 0.9996 & 0.1056 & 0.0886 & 0.0982 & 0.9996 \\
   & Swish & \( x \cdot \frac{a}{1+e^{-ax}} \) & 0.9994 & 0.0507 & 0.0398 & 0.0560 & 0.1594 \\
    \midrule
BGD & Swish & \( x \cdot \frac{a}{1+e^{-ax}} \) & 0.9998 & 0.8469 & 0.9581 & 0.9857 & 0.9996 \\
  \bottomrule[1pt]
  \multicolumn{8}{l}{*The parameter a of activation functions is used with the default value of 1.} \\
\end{tabular}
\end{table*}

\section{Experimental Results}\label{sec:experimental-results}
In this section, we evaluate the representational capacity of the designed FF-PINNs architecture in modeling the propagation of acoustic waves through complex infinite media. We observed that PINNs without Fourier feature exhibit relative $\ell_2$ errors close to 1 for all cases in this section. This indicates that vanilla PINNs cannot learn the phenomenon of high-frequency wave propagation around a point source. In contrast, FF-PINNs exhibit satisfactory performance. Before conducting FF-PINN simulations, we construct a velocity surrogate model using coordinate-velocity label data. We train Gaussian Fourier feature NNs with 5 hidden layers, each containing 20 neurons, employing an Adam optimizer with a learning rate of $5 \times 10^{-3}$ for 100,000 epochs, where $\sigma=15$. All the numerical implementations are coded in Jax \cite{Bradbury_2018_JAXComposableTransformations} and performed on an NVIDIA Tesla V100 GPU card (32G) in a standard workstation.

In this work, we use the self-adaptive weighting scheme based on Neural Tangent Kernel (NTK) theory \cite{Wang_2022_WhenWhyPINNs} to determine appropriate weights $\lambda_r$ and $\lambda_{ce}$ during training. See the Appendix \ref{sec:adapt-weight-param} for a detailed discussion of the weighting parameters. The accuracy impact due to ABCs is shown in Table \ref{tab:3}, where the number of the ABCs collocation points sampled randomly from every boundary in each training iteration is $N_{ce}=2000$. The term "w/ ABCs" in Table \ref{tab:3} refers to the second-order Clayton-Engquist acoustic paraxial approximation boundary applied on all four boundaries of the rectangular domain. The computational model settings and neural network hyperparameters are summarised in Table \ref{tab:4}.

\begin{table}[!t]
  \caption{Comparison of the relative $\ell_2$ errors for different cases with and without ABCs}
  \label{tab:3}
  \centering
  \setlength\tabcolsep{0.1pt}
  \begin{tabular}{m{1.5cm}<{\centering} m{1.5cm}<{\centering} m{2.5cm}<{\centering} m{1.5cm}<{\centering} m{1.5cm}<{\centering}}
  \toprule
  Boundary Setting & 4-layer & Random mixed Gaussian velocity & Marmousi model & Overthrust model \\
    \midrule
    w/ ABCs & 0.0582 & 0.0613 & 0.0751 & 0.0816 \\
    w/o ABCs & 0.0763 & 0.0976 & 0.1806 & 0.2128 \\
    \bottomrule \\
\end{tabular}
\end{table}

\begin{table*}[!t]
  \caption{Computational model settings and neural network hyperparameters}
  \label{tab:4}
\centering
\begin{tabular}{cccccc}
  \toprule[1pt]
  Cases & Network ([width]$\times$depth) & Collocations & $f_0$ (Hz) & $\alpha$ & $\sigma$ \\
  \midrule
  4-layer & [80] $\times$ 5 & $N_r = 80,000$ & 5 & 0.01 & 1 \\
  Random mixed Gaussian velocity & [80] $\times$ 8 & $N_r = 80,000$ & 10 & $\{0.02, 0.03, 0.025\}$ & 1 \\
  Marmousi model & [128] $\times$ 5 & $N_r = 80,000$ & 10 & 0.01 & 1.5 \\
  Overthrust model & [128] $\times$ 5 & $N_r = 80,000$ & 15 & 0.02 & 1.5 \\
  \bottomrule[1pt]
\end{tabular}
\end{table*}

\subsection{4-layer velocity model}
\begin{figure}[!t]
\centering
\includegraphics[width=3.2in]{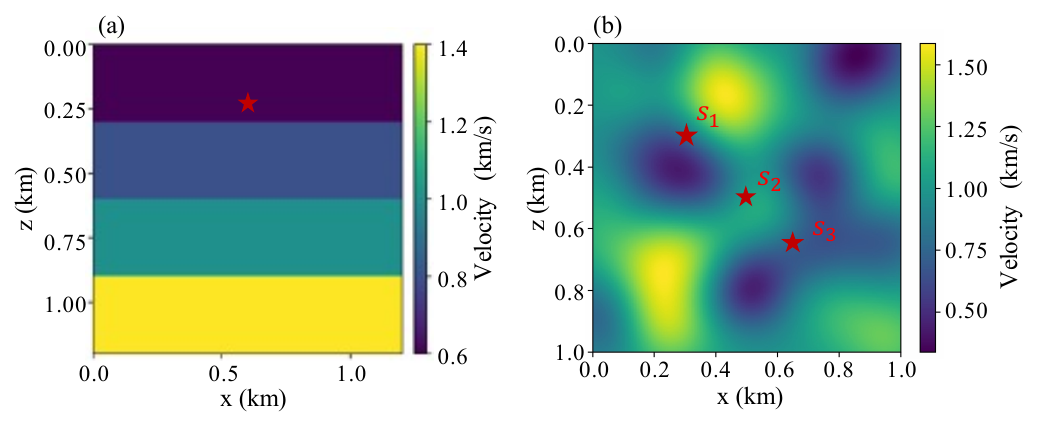}
\caption{(a) 4-layer velocity model. (b) Random mixed Gaussian velocity model.}
\label{fig_4layers_Gaussian_models}
\end{figure}

A 4-layer velocity model case is designed to assess the ability of FF-PINNs to learn the wavefield solutions in horizontal layered media. As shown in Fig. \ref{fig_4layers_Gaussian_models}a, the computational model is a rectangular domain with a width of 1.2 km and a height of 1.2 km. We have set up four layers distributed evenly along the depth, with velocities of 0.6 km/s, 0.8 km/s, 1.0 km/s, and 1.4 km/s respectively. A Ricker source excitation with $f_0=5$ Hz and $\alpha=\ 0.01$ is applied at $\left\{x_s,z_s\right\}=\left\{0.6,\ 0.24\right\}$ km in the spatial domain. The total calculated time for wave propagation is 1.2 s.

\begin{figure}[!t]
\centering
\includegraphics[width=3.2in]{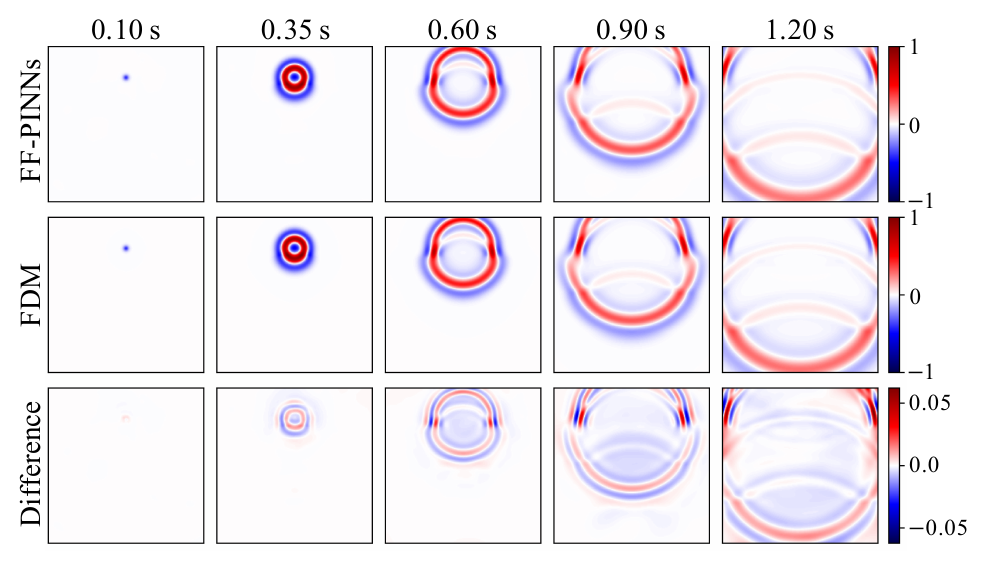}
\caption{Comparison of wavefields of FF-PINNs and FDM in the 4-layer velocity model case.}
\label{fig_4layers_wavefield}
\end{figure}

The FF-PINNs contain 5 hidden layers, 80 neurons per layer, and a standard deviation $\sigma=1$. The PDE residual collocation points with $N_r=80,000$ were randomly sampled from the spatio-temporal domain in each training iteration. Fig. \ref{fig_4layers_wavefield} illustrates the comparison between the results from the staggered-grid finite difference method and FF-PINNs with ABCs, after 30,000 training epochs using the Adam optimizer. Seismic waves propagating through layered velocity interfaces give rise to reflection and transmission phenomena, as well as compression and expansion of the waves. Under the same experimental setup, only considering the effect of ABCs soft regularization constraint. The relative $\ell_2$ error of the FF-PINNs prediction with ABCs is 0.0582, while the relative $\ell_2$ error without ABCs is 0.0763. Since the wave velocity model is not complex, FF-PINNs without considering ABCs can also learn the behavior of transmitted outgoing waves. Moreover, the additional ABCs loss term may slow down the optimization of the PDEs loss term in the early training epochs, which can explain the small difference in the errors. Upon examining the wavefield snapshots presented in Fig. \ref{fig_4layers_wavefield}, it becomes evident that the FF-PINNs adeptly capture distinct physical phenomena occurring near the medium interfaces.

\subsection{Random mixed Gaussian velocity model}
We further increase the complexity of the wavefield by using three source terms with different $\alpha$ values and considering a random mixed Gaussian velocity model. Based on a background wave velocity of $c_0\left(\mathbf{x}\right)=1.0$ km/s, the following random mixed Gaussian model is incorporated
\begin{equation}
G_{c}(\mathbf{x})=M_{c} \exp \left(-\frac{1}{2}\left\|\frac{\mathbf{x}-\mathbf{x}_{c}}{\alpha_{c}}\right\|_{2}^{2}\right),
\end{equation}
where $M_c$ represents the amplitude, which is set as a random number between $\left[-3,\ 3\right]$. The location $\mathbf{x}_c$ is uniformly distributed between $\left[0,\ 1\right]$. The $\alpha_c$ is set to 1. The resulting random mixed Gaussian velocity model ($c_0\left(\mathbf{x}\right)+G_c\left(\mathbf{x}\right)$) is illustrated in Fig. \ref{fig_4layers_Gaussian_models}b.

\begin{figure}[!t]
\centering
\includegraphics[width=3.2in]{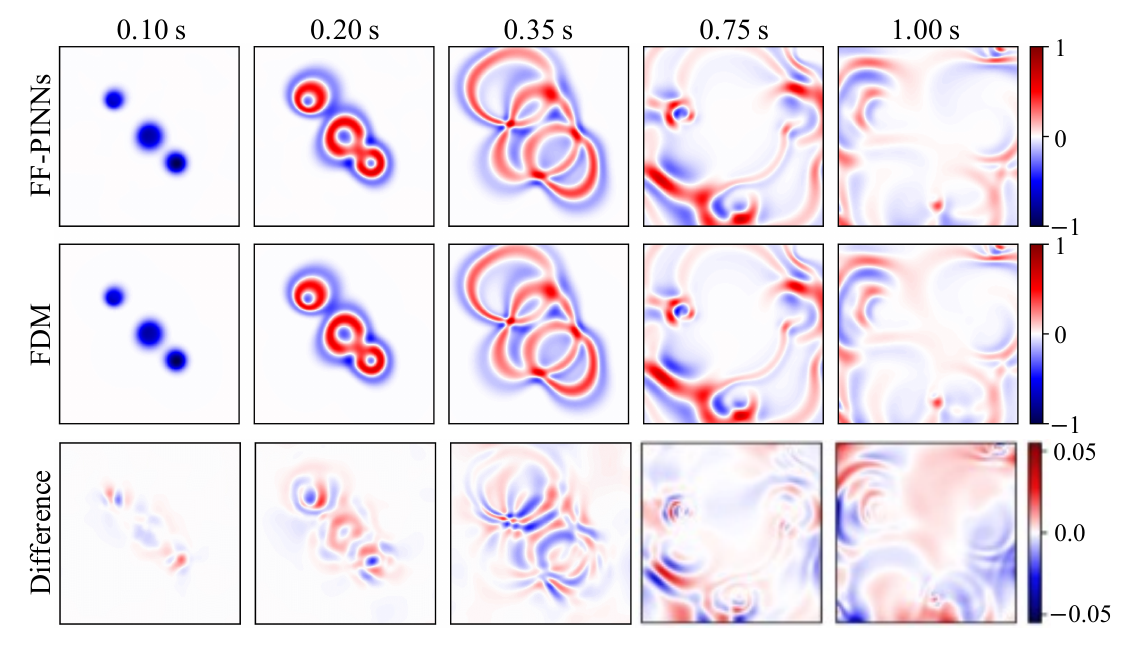}
\caption{Comparison of wavefields of FF-PINNs and FDM in the random mixed Gaussian velocity model case.}
\label{fig_Gaussian_Mixed_wavefield}
\end{figure}

We consider the Ricker wavelet time function with a duration of 1.0 s and a central frequency of 10 Hz. The kernel width $\alpha$ of the three sources $\left\{s_1,s_2,s_3\right\}$ are 0.02, 0.03, and 0.025, and the source locations $\mathbf{x}_c=\left\{x_c,z_c\right\}$ are $\left\{0.3,0.3\right\}$ km, $\left\{0.5,\ 0.5\right\}$ km and $\left\{0.65,\ 0.65\right\}$ km, respectively. The FF-PINNs with standard deviation of $\sigma=1$ is used to approximate the displacement solution. The neural networks consist of 8 hidden layers with 80 neurons per layer. During each training iteration, the number of PDE residual collocation points randomly sampled from the spatio-temporal domain is $N_r=80,000$. After 30,000 epochs of training using the Adam optimizer, the relative $\ell_2$ errors of the predicted solutions for FF-PINNs with and without ABCs are 0.0613 and 0.0976, respectively. Table \ref{tab:3} shows the corresponding comparison. Fig. \ref{fig_Gaussian_Mixed_wavefield} shows the comparison between the FF-PINNs prediction with ABCs and the FDM results. It is clear that FF-PINNs learn the complex wave propagation dynamics generated by the three sources, and the predicted solution matches the ground truth very well.

\subsection{2D Marmousi model and Overthrust model}
\begin{figure}[!t]
\centering
\includegraphics[width=3.2in]{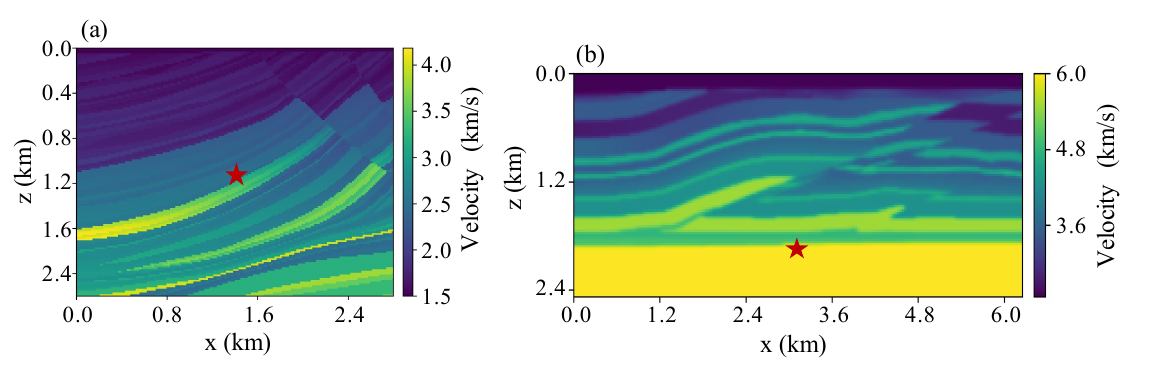}
\caption{Computational model. (a) 2D Marmousi model. (b) 2D Overthrust model.}
\label{fig_Mar_Over_model}
\end{figure}

Next, we apply the proposed framework to the extracted Marmousi model and Overthrust model, as shown in Fig. \ref{fig_Mar_Over_model}. To test the capability of FF-PINNs in handling velocity interfaces with sharp variations, we did not apply any smoothing to the velocity models. For the extracted Marmousi model, the model size is $176 \times 138$, and the spatial sampling interval is 16 m in both $x$ and $z$ directions. The source time function is the Ricker wavelet at the center of the model with $f_0=10$ Hz and $\alpha=\ 0.01$ located. For the extracted Overthrust model, the model size is $201\times100$ and the spatial sampling interval is 25 m. The center frequency $f_0$ of the Ricker wavelet is 15 Hz and $\alpha=0.02$. The total computational time for wave propagation in both cases is 1.0 second.

\begin{figure*}[!t]
\centering
\includegraphics[width=4.5in]{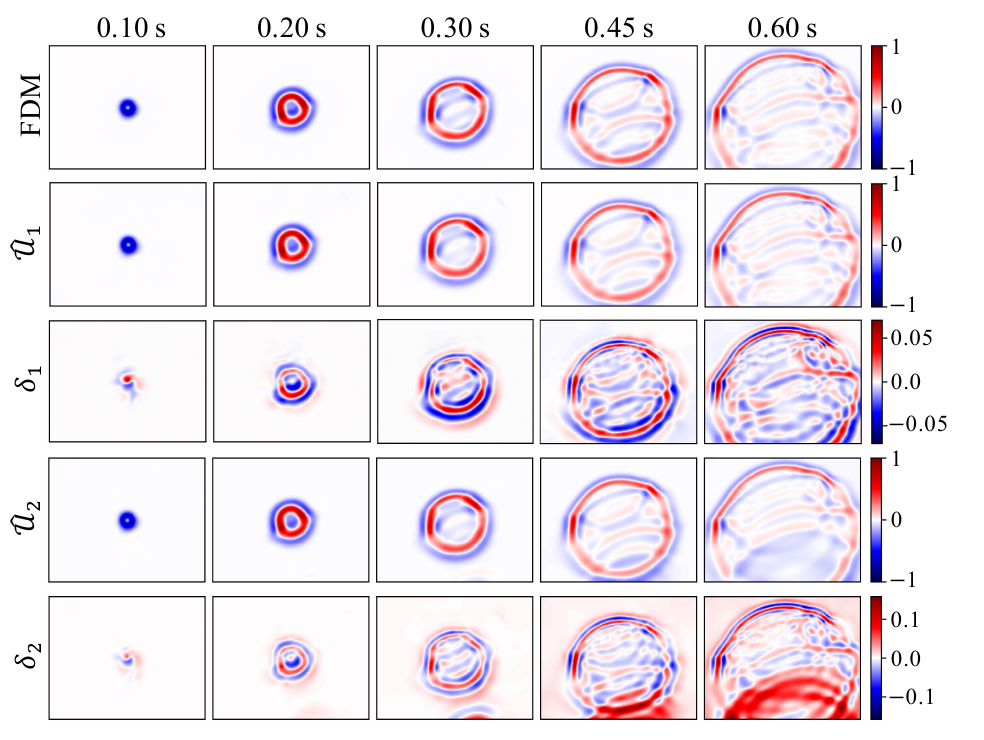}
\caption{Comparison of wavefields of FF-PINNs and FDM in the Marmousi model case. Predicted wavefields for FF-PINNs (${\hat{\mathcal{U}}}_1$: with ABCs; ${\hat{\mathcal{U}}}_2$: without ABCs). The absolute error of the predicted wavefield by FF-PINNs ($\delta_1$: with ABCs; $\delta_2$: without ABCs).}
\label{fig_Mar_wavefield}
\end{figure*}

The FF-PINNs with standard deviation of $\sigma=1.5$ is used to approximate the displacement solution. The neural networks consist of 5 hidden layers with 128 neurons per layer. The number of PDE residual collocation points randomly sampled from the spatio-temporal domain is $N_r=80,000$.

Due to the increased complexity of the Marmousi and Overthrust models under consideration, the sequential training strategy of time-domain decomposition introduced in Section \ref{sec:experimental-setup} is applied to both of these cases. This training strategy is consistent with the laws of physics for the time-dependent dynamical systems. We initially conducted 5,000 pre-training epochs using the Adam optimizer within the time range of $[0, 0.3]$ s, followed by 10,000 epochs of training within the time ranges of $[0, 0.4]$ s and $[0, 0.5]$ s, and finally, 20,000 epochs within the range of $[0, 0.6]$ s to obtain the solution for the entire time domain.

\begin{figure*}[!t]
\centering
\includegraphics[width=4.5in]{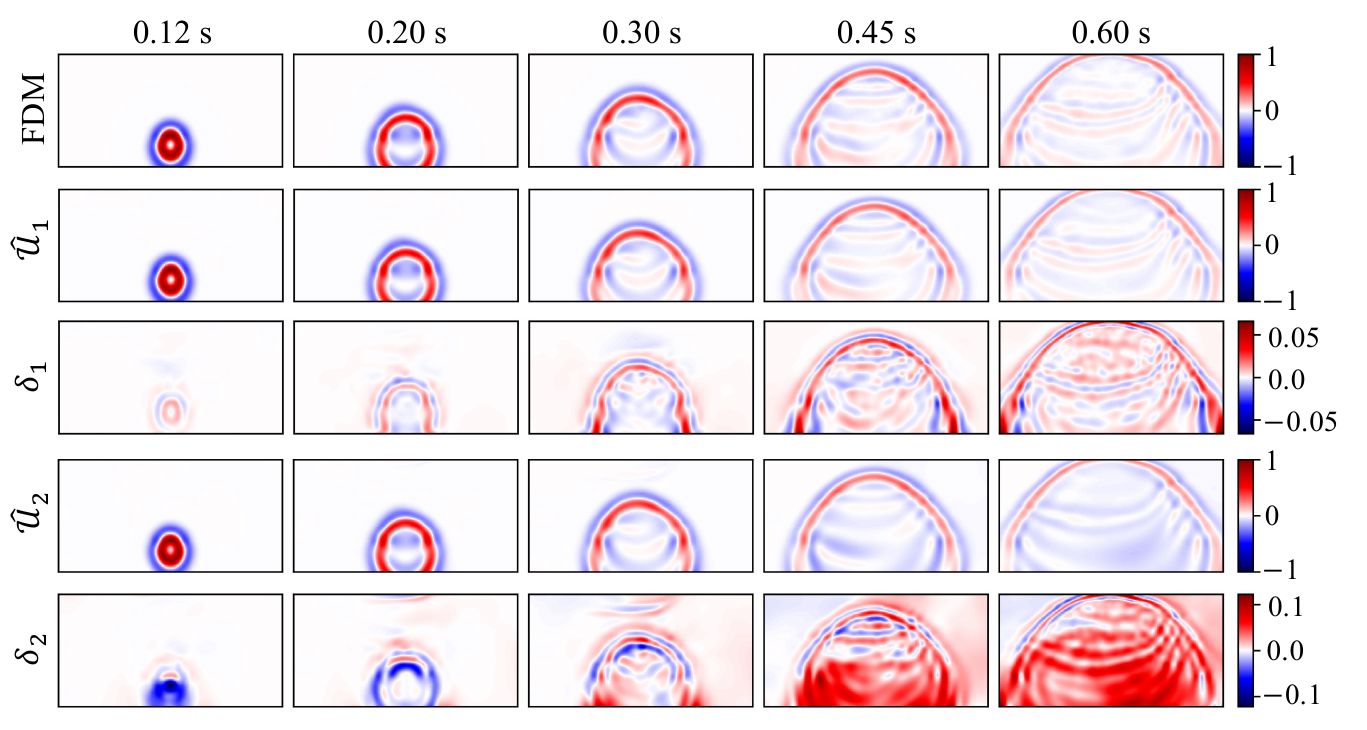}
\caption{Comparison of wavefields of FF-PINNs and FDM in the Overthrust model case. Predicted wavefields for FF-PINNs (${\hat{\mathcal{U}}}_1$: with ABCs; ${\hat{\mathcal{U}}}_2$: without ABCs). The absolute error of the predicted wavefield by FF-PINNs ($\delta_1$: with ABCs; $\delta_2$: without ABCs).}
\label{fig_Over_wavefield}
\end{figure*}

We show in Table \ref{tab:3} the accuracy comparison between the Marmousi case and the Overthrust case with and without ABCs. With ABCs, the relative $\ell_2$ errors are 0.0751 and 0.0816 for the two cases, respectively. However, without ABCs, the errors significantly increase to 0.1806 and 0.2128, respectively. Figs. \ref{fig_Mar_wavefield} and \ref{fig_Over_wavefield} show the snapshots of the predictions of FF-PINNs (with and without ABCs) compared to the FDM-simulated wavefields for the 5 moments of the Marmousi and Overthrust models. Compared to the incident waves, the refracted and reflected waves are much smaller in scale. This poses a significant challenge for fully-connected neural networks as approximators in PINNs. As can be observed, the continuous fine-grained time-domain decomposition training strategy and the soft regularization of ABCs have enabled FF-PINNs to capture the details of waves, yielding high-fidelity wavefield simulation results. When the ABCs soft regularization constraint is not imposed, spurious reflected waves are found to be generated around the boundaries. This leads to very large errors near the boundaries, which further extend to the full spatial domain, severely affecting the accuracy performance of FF-PINNs. On the contrary, the proper application of ABCs can effectively attenuate or even eliminate unreasonable errors at the boundaries, leading to a clear and accurate wavefield solution. This provides strong evidence supporting the necessity of imposing ABCs within the proposed framework.

\section{Discussion}
\textbf{Boundary conditions}. This work focuses on the necessity and applicability of ABCs in solving 2D acoustic equations in infinite complex media using FF-PINNs. Free surface boundary conditions are not imposed in this work, mainly because of the possible failure of soft-constrained zero-stress boundary conditions \cite{Ding_2023_PhysicsconstrainedNeuralNetworks}, which is still an open problem to be solved by the community. Hard-embedded boundary conditions are a reasonable and promising approach. For second-order wave equations, hard embedding of homogeneous Dirichlet boundary conditions (fixed boundaries) is straightforward for PINNs. However, hard embedding of the homogeneous Neumann boundary conditions cannot be achieved by directly modifying the neural network output. Developments in mesh-free methods and the deep learning community provide theoretical support for potential implementations \cite{Lyu_2021_EnforcingExactBoundary, Sukumar_2022_ExactImpositionBoundary}.

\textbf{Efficacy of sharp velocity models}. When dealing with non-smooth complex media, PINNs will tend to give smooth wavefields. This may be attributed to two reasons: 1) “spectral bias” pathologies that make learning high frequencies challenging, and 2) commonly used activation functions, such as Tanh, tend to produce smooth output results \cite{Sitzmann_2020_ImplicitNeuralRepresentations}. The FF-PINNs and Swish activation function used in this work significantly enhanced the efficacy and accuracy of neural networks in handling sharp velocity models. It is worth mentioning that updating the randomly sampled collocation points (SGD strategy and velocity surrogate model) at each training epoch is crucial for mitigating point source singularity and learning the details of small scattered and reflected waves. Thus, the proposed approach is an extensible, unified framework with high resolution and reasonable accuracy. Our future work includes validating model performance in more challenging exploration settings, such as those with irregular topography, high heterogeneity or fault structures.

\textbf{Computational efficiency}. The second-order ABCs in our proposed framework involve additional second-order derivative calculations, which will increase the training cost to some extent. Nevertheless, the previous numerical case comparisons have provided evidence for the necessity of imposing ABCs. Recent studies have added source location as an additional input parameter to the network \cite{Song_2022_SimulatingSeismicMultifrequency, Ren_2024_SeismicNetPhysicsinformedNeural} and explored transfer learning \cite{Wu_2023_HelmholtzequationSolutionNonsmooth} to improve simulation efficiency. However, computational efficiency remains a major limitation for the application of PINNs methods to large-scale computational scenarios. PINNs method requires the evaluation of PDEs residuals over a large number of sampling points on the internal domain, and the computation of higher-order derivatives through automatic differentiation can be time-consuming. Subdomain methods \cite{Kharazmi_2021_HpVPINNsVariationalPhysicsinformed, Diao_2023_SolvingMultimaterialProblems, Dolean_2024_MultilevelDomainDecompositionbased}, discrete learning methods \cite{Ren_2022_PhyCRNetPhysicsinformedConvolutionalrecurrent, Rao_2023_EncodingPhysicsLearn}, and neural spectral methods \cite{Du_2023_NeuralSpectralMethods} have the potential to reduce the computational burden of learning seismic wave propagation on a large scale. In addition, using the linearly shrinking-type NNs \cite{Chai_2024_PracticalAspectsPhysicsInformed, Wu_2024_PINNBasedSeismicWavefield} instead of tube-type NNs and employing a neuron splitting method \cite{Huang_2022_PINNupRobustNeural} in the time domain decomposition strategy, has the potential to further enhance the computational efficiency of the proposed framework. With these more efficient learning frameworks, we believe that the proposed framework for solving Fourier-featured PINNs may provide a viable solution for 3D modelling in large complex media.

\textbf{Better frequency manipulation}. The conventional initialization method for Fourier feature NNs based on empirical or hyperparameter sweep should still be used, which is a major shortcoming of this work. If observations are available, the initial values of the Fourier feature parameters can be manipulated based on the spectrum information of the target function. Furthermore, our framework embeds the spatio-temporal coordinates into the high frequency space with the same scale factor $\sigma$. However, the "frequencies" in the temporal and spatial domains may be different, or even the gap between them may be significant. In such cases, spatio-temporal multi-scale Fourier feature embedding \cite{Wang_2021_EigenvectorBiasFourier} may be a promising idea. The frequency of the Fourier feature mapping determines the frequency of the NTK feature vectors, which can provide a more fundamental guidance of the training dynamics through the perspective of the NTK theory \cite{Zeng_2024_TrainingDynamicsPhysicsInformed}. In conclusion, how to rigorously establish a general theory of the relationship between frequency and Fourier features is an open problem that requires theoretical support from more future research.

\textbf{Integration with data}. One of the advantages of PINNs is the seamless integration of physics and data. However, real-world seismic exploration often involves more complex geological scenarios. Observations are typically sparse and noisy, which presents challenges for integrating data with physical models. To what extent can real-world observational data improve forward modeling? We believe this is an intriguing and meaningful direction for future research, and we hope to provide answers in due course. The vanilla PINNs used by Rasht-Behesht et al. (2022) \cite{Rasht-Behesht_2022_PhysicsInformedNeuralNetworks} struggles to identify all the internally reflected phases at sharp material discontinuities, leading to poor inversion results. We suggest that this may be due to a spectral bias in the forward modelling of vanilla PINNs. The use of sensible Fourier feature mappings can enhance the ability to learn details of the reflected waves, thus potentially attenuating this smoothness. The velocity surrogate model proposed in this work is essentially a reparameterization of the subsurface velocity model using a Fourier feature NNs. However, the effectiveness of reparametrizing the velocity model with Fourier feature NNs in FWI remains to be investigated. Using higher values of $\sigma$ tends to introduce salt-and-pepper artifacts that are susceptible to noise. The use of varying $\sigma$ at different stages of training may be a viable option and will be further investigated in future work.

\section{Conclusion}
In this study, we present a unified framework of Fourier feature PINNs for modeling high-frequency wave propagation in sharp and complex media. The accuracy comparison of the activation functions and sampling strategies in dealing with the point source problem is discussed in detail. The combination of the Swish activation function and the SGD strategy achieves the best performance in all the experiments. The independently pre-trained wave velocity surrogate model provides fast and accurate wave velocity predictions in each training epoch. Fourier feature neural networks sampled from different families of distributions (Gaussian, Laplace, and uniform) have similar error distribution patterns. The standard deviation of the distribution family determines the range of coordinate frequency mapping and is an important parameter affecting the accuracy of FF-PINNs. For the case of non-smooth complex media, the imposition of ABCs avoids spurious boundary-reflected waves and significantly improves the ability of NNs to learn detailed waves. Realistic assessments demonstrate the efficacy of the proposed framework compared to the vanilla PINNs, particularly in scenarios involving high-frequencies and non-smooth, heterogeneous real-world models.

\section*{Acknowledgments}
The work reported here is partially supported by research grants from the National Science Foundation of China (NSFC, Grant Nos. 52192675, U1839202). The authors thank the editor and the reviewers for constructive comments on improving the paper.

% \appendix
\appendices
\section{The details of weighting hyper-parameters}\label{sec:adapt-weight-param}
One of the main challenges in training PINNs is addressing multi-scale losses that arise from the minimization of PDE residuals. In the experiments, the loss function optimization involves balancing different physical constraints, such as PDE and boundary conditions. Taking the 4-layer and Marmousi model as examples, we fixed $\lambda_r=1$ and varied the values of $\lambda_{ce}$ to conduct a sensitivity analysis of loss weights. Table \ref{tab:5} demonstrates the results of the sensitivity analysis. $\lambda_{ce}$ controls the effect of the ABCs. As $\lambda_{ce}$ decreases, it is closer to not applying ABCs. For cases where the wavefield near the boundary is not complex (e.g., in the 4-layer case), this effect is not significant. However, for velocity model with complex sharp interfaces near the boundary (e.g., Marmousi model), a large error arises. On the other hand, a large value of $\lambda_{ce}$ affects the optimization of the PDE loss term, again resulting in a large error.

\begin{figure}[!t]
\centering
\includegraphics[width=3.0in]{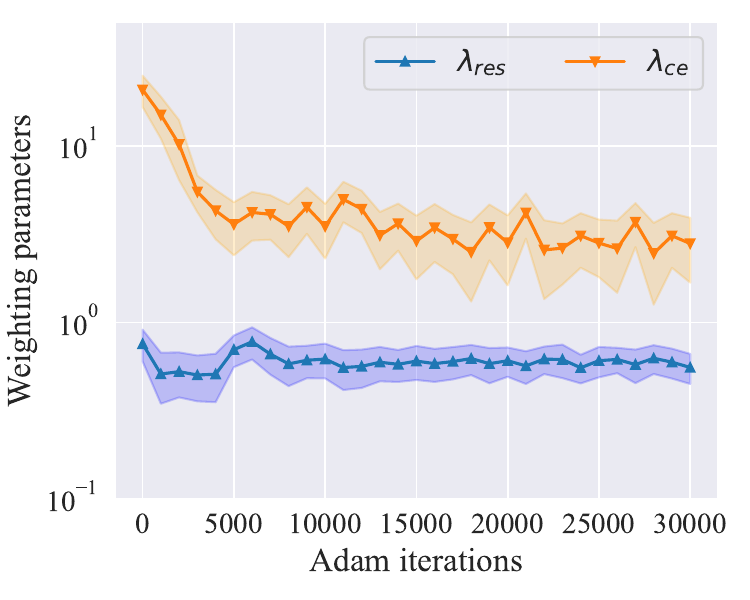}
\caption{4-layer case: the evolution of the weight parameters using the NTK adaptive weighting algorithm over 5 different random seeds. The shaded region shows the 2nd to 98th percentile of aleatory variability.}
\label{fig_AdaptiveParams}
\end{figure}

Wang et al. (2022) proposed a novel gradient descent algorithm through the lens of the NTK theory to adaptively calibrate the convergence rate of the total training error \cite{Wang_2022_WhenWhyPINNs}. This scheme determines the weights of the various loss terms based on the sum of the eigenvalues of the component NTK matrix (equivalent to the trace of the matrix). In the experiments, we use the NTK-based loss weighting scheme to determine appropriate weights $\lambda_r$ and $\lambda_{ce}$ during training. The NTK of the PDE residual term is calculated by randomly sampling 500 points in the spatio-temporal domain, with 100 sampling points for each of the four boundaries. We update loss weights every 1,000 Adam iterations by default. Figure \ref{fig_AdaptiveParams} shows the evolution of the weighting parameters for the 4-layer case using the NTK-based scheme. The result is obtained over 5 different random seeds. The relative $\ell_2$ error using the NTK-based weighting scheme is shown in Table \ref{tab:3}. By comparing the results in Table \ref{tab:3} and Table \ref{tab:5}, it can be observed that the NTK-based weighting scheme automatically determines reasonable weights, avoiding manual empirical tuning, which could lead to significant computational overhead.

\begin{table*}[!t]
  \caption{Sensitivity analysis of loss weights. Comparison of relative $\ell_2$ errors caused by different $\lambda_{ce}$ when $\lambda_r$ is fixed at 1.}
  \label{tab:5}
  \centering
\begin{tabular}{cccccccc}
  \toprule
  \multirow{2}{*}{  Cases } & \multicolumn{7}{c}{ $\lambda_{ce}$ } \\
\cmidrule { 2 - 8 } & 0.001 & 0.01 & 0.1 & 1 & 5 & 10 & 100 \\
  \midrule
  4-layer & 0.0728 & 0.0891 & 0.0656 & 0.0614 & 0.0746 & 0.0895 & 0.2039 \\
   Marmousi model & 0.1455 & 0.1136 & 0.0965 & 0.0787 & 0.1063 & 0.1454 & 0.2902 \\
    \bottomrule \\
\end{tabular}
\end{table*}

\bibliographystyle{IEEEtran}
\bibliography{zotero_betterBib_shortJournalTitle}

\end{document}